# PUBLIC KEY CRYPTOGRAPHY BASED ON TWISTED DIHEDRAL GROUP ALGEBRAS

Javier de la Cruz

Department of Mathematics and Statistics
Universidad del Norte
Barranquilla, Colombia

Ricardo Villanueva-Polanco*

Department of Computer Science and Engineering
Universidad del Norte
Barranquilla, Colombia

ABSTRACT. In this paper, we propose to use a twisted dihedral group algebra for public-key cryptography. For this, we introduce a new 2-cocycle $\alpha_\lambda$ to twist the dihedral group algebra. Using the ambient space $\mathbb{F}^{\alpha_\lambda} D_{2n}$, we then introduce a key exchange protocol and present an analysis of its security. Moreover, we explore the properties of the resulting twisted algebra $\mathbb{F}^{\alpha_\lambda} D_{2n}$, exploiting them to enhance our key exchange protocol. We also introduce a probabilistic public-key scheme derived from our key-exchange protocol and obtain a key encapsulation mechanism (KEM) by applying a well-known generic transformation to our public-key scheme. Finally, we present a proof-of-concept implementation of the resulting key encapsulation mechanism.

1. **Introduction.** With the advent of building a large quantum computer in years ahead, most of the current public-key schemes will be rendered insecure. Thus it is important to search for alternative approaches to the current public-key schemes. In this quest for these alternatives, several candidates have already been proposed, which are believed to be resistant to an adversary with access to a sufficiently large quantum computer, and are based on various families of mathematical problems. In particular, these families may be categorized into lattice-based cryptography, multivariate cryptography, hash-based cryptography, code-based cryptography, and supersingular elliptic curve isogeny cryptography. Additionally, the National Institute of Standards and Technology (NIST) is running a post-quantum cryptography standardization process whose goal is to standardize a set of public-key schemes for future use [18].

Recent works have explored some constructions of codes and cryptographic constructions over twisted group algebras [5, 10]. On the one hand, the authors in [5] investigate right ideals as codes in twisted group algebras. In particular they characterize all linear codes which are twisted group codes in terms of their automorphism group. On the other hand, the authors in [10] propose public key protocols over twisted dihedral group rings. In particular, they propose a key exchange protocol in the style of Diffie-Hellman over twisted dihedral group rings and a computationally equivalent cryptosystem. These cryptographic constructions seem to be quantum-resistant.

---

* Corresponding author: Ricardo Villanueva-Polanco (rpolanco@uninorte.edu.co).

In this paper, we explore an alternative approach and propose a key exchange protocol in the style of Diffie-Hellman based on a twisted dihedral group algebra. First, we introduce a new 2-cocycle $\alpha_\lambda$ to twist an dihedral group algebra. This 2-cocycle is connected to the algebraic properties of the anti-isomorphism of $\mathbb{F}$ algebras $\widehat{\phantom{x}}: \mathbb{F}^{\alpha_\lambda} G \longrightarrow \mathbb{F}^{\alpha_\lambda^{-1}} G$, which helps in realizing our proposed protocol and its computations. Second, we explore the properties of the resulting twisted algebra $\mathbb{F}^{\alpha_\lambda} G$, exploiting them to enhance our key exchange protocol. We then compare our protocol against a recent proposal [10] that also uses a twisted dihedral group algebra. We prove that each protocol operates on a structurally different twisted group algebra, i.e. they are different. We then prove our construction is session-key secure in the authenticated-links adversarial model of Canetti and Krawczyk [3], assuming the Decisional Dihedral Product (DDP) Assumption holds for $\mathbb{F}^{\alpha_\lambda} D_{2n}$. Additionally, we construct a probabilistic public-key scheme derived from our key exchange protocol and prove it is CPA-secure, assuming the DDP Assumption holds for $\mathbb{F}^{\alpha_\lambda} D_{2n}$. We finally obtain a key encapsulation mechanism by applying a well-known generic transformation to our public-key scheme [12].

This paper is structured as follows. In Section 2 we collect preliminaries on 2-cocycles and twisted group algebras. In Section 3 we analyze the structure of a general twisted group algebra when the group is $D_{2n}$, the dihedral group of order $2n$. Moreover, we define a particular 2-cocycle $\alpha_\lambda$ for this group $D_{2n}$, which we will use to construct the twisted dihedral group algebra $\mathbb{F}^{\alpha_\lambda} D_{2n}$. In Section 4, we introduce our key exchange protocol and present an analysis of its security. In Section 5, we introduce a probabilistic public-key scheme derived from the key exchange protocol and also present its corresponding security analysis. In section 6, we present a key encapsulation mechanism obtained from applying a generic transformation to our probabilistic public-key scheme and present details of a proof-of-concept implementation. Finally, in Section 7 we conclude our work and present future research works.

2. **Preliminaries.** Throughout $\mathbb{F}$ will be the finite field with $q$ elements and $G$ a finite multiplicative group. We summarise the definitions and properties we need on twisted group algebras. We call the map $\alpha : G \times G \longrightarrow \mathbb{F}^* = \mathbb{F} \setminus \{0\}$ a *2-cocycle* of $G$ if $\alpha(1,1) = 1$ and for all $g, h, k \in G$ it satisfies the equation $\alpha(g, hk)\alpha(h, k) = \alpha(gh, k)\alpha(g, h)$. We denote the set of all 2-cocycles of $G$ by $Z^2(G, \mathbb{F}^*)$. For $\alpha, \beta \in Z^2(G, \mathbb{F}^*)$, we define $\alpha\beta \in Z^2(G, \mathbb{F}^*)$ by $\alpha\beta(g, h) = \alpha(g, h)\beta(g, h)$ for all $g, h \in G$. With this operation $Z^2(G, \mathbb{F}^*)$ becomes a multiplicative abelian group. If $\beta : G \longrightarrow \mathbb{F}^*$ is a map such that $\beta(1) = 1$, the *coboundary* $\partial\beta$ defined by $\partial\beta(g, h) = \beta(g)^{-1}\beta(h)^{-1}\beta(gh)$ for all $g, h \in G$ is in $Z^2(G, \mathbb{F}^*)$. Let $B^2(G, \mathbb{F}^*)$ denote the set of all coboundaries of $G$. The set $B^2(G, \mathbb{F}^*)$ of coboundaries forms a subgroup of the group $Z^2(G, \mathbb{F}^*)$ and for $\alpha \in Z^2(G, \mathbb{F}^*)$ we denote its coset by $[\alpha] := \alpha B^2(G, \mathbb{F}^*)$. The factor group $H^2(G, \mathbb{F}^*) = Z^2(G, \mathbb{F}^*)/B^2(G, \mathbb{F}^*) = \{[\alpha] : \alpha \in Z^2(G, \mathbb{F}^*)\}$ is usually called the second cohomology group of $G$ with values in $\mathbb{F}^*$.

**Definition 2.1.** Let $\alpha$ be a 2-cocycle of $G$. The twisted group algebra $\mathbb{F}^\alpha G$ is the set of all formal sums $\sum_{g \in G} a_g \bar{g}$, where $a_g \in \mathbb{F}$, with the following twisted

multiplication
$$\overline{g}\overline{h} = \alpha(g,h)\overline{gh}.$$

It is well known that if $\alpha, \beta \in Z^2(G, \mathbb{F}^*)$, then $\mathbb{F}^\alpha G \cong \mathbb{F}^\beta G$ as algebras if and only if $[\alpha] = [\beta]$ ([15], Proposition 1.2.6).

**Lemma 2.2.** *For any $\alpha \in Z^2(G, \mathbb{F}^*)$ the map $\varphi : \mathbb{F}^\alpha G \longrightarrow \mathbb{F}^{\alpha^{-1}} G$, $\sum_{g \in G} a_g \overline{g} \longmapsto \sum_{g \in G} a_g \overline{g}^{-1}$, where $\overline{g}^{-1} = \alpha(g, g^{-1})\overline{g^{-1}}$ is the inverse of $\overline{g}$ in $\mathbb{F}^{\alpha^{-1}} G$, is an anti-isomorphism of $\mathbb{F}^\alpha G$ onto $\mathbb{F}^{\alpha^{-1}} G$.*

*Proof.* Let $g, h \in G$. Then we have
$$\varphi(\overline{g}\overline{h}) = \varphi(\alpha(g,h)\overline{gh}) = \alpha(g,h)(\overline{gh})^{-1} = \alpha(g,h)(\alpha(g,h)\overline{gh})^{-1} = \overline{h}^{-1}\overline{g}^{-1} = \varphi(\overline{h})\varphi(\overline{g}).$$
$\square$

**Definition 2.3.** [5] For an element $a = \sum_{g \in G} a_g g \in \mathbb{F}^\alpha G$ we define its adjunct as
$$\widehat{a} := \varphi(a) = \sum_{g \in G} a_g \alpha(g, g^{-1}) \overline{g^{-1}}.$$

**Remark 1.** Note that if $\alpha = \alpha^{-1}$, then by Lemma 2.2 the map $a \mapsto \widehat{a}$ defines an algebra anti-isomorphism of $\mathbb{F}^\alpha G$ of order 2.

3. **A twisted dihedral group algebra.** In this section, we let $G = D_{2n}$, where $D_{2n} = \langle x, y : x^n = y^2 = 1, yxy^{-1} = x^{-1} \rangle$ is the dihedral group of order $2n$.

**Notation.** In what follows the symbol $[k]_n$ for $k \in \mathbb{Z}$ denotes $k \equiv [k]_n \mod n$.

**Lemma 3.1.** *Let $C_n = \langle x \rangle$ be the cyclic subgroup of $D_{2n}$ generated by $x$ and $\alpha$ a 2-cocycle of $D_{2n}$. Then we have*
  1. *$\mathbb{F}^\alpha D_{2n}$ is a free $\mathbb{F}^\alpha C_n$-module with basis $\{1, y\}$. Therefore $\mathbb{F}^\alpha D_{2n} = \mathbb{F}^\alpha C_n \oplus \mathbb{F}^\alpha C_n y$ as direct sum of $\mathbb{F}$-vector spaces.*
  2. *$\mathbb{F}^\alpha C_n y \cong \mathbb{F}^\alpha C_n$ as $\mathbb{F}^\alpha C_n$-modules.*
  3. *For $a \in \mathbb{F}^\alpha C_n y$, $ab \in \mathbb{F}^\alpha C_n$ if $b \in \mathbb{F}^\alpha C_n y$ or $ab \in \mathbb{F}^\alpha C_n y$ if $b \in \mathbb{F}^\alpha C_n$.*
  4. *If $a \in \mathbb{F}^\alpha C_n$, then $\widehat{a} \in \mathbb{F}^\alpha C_n$.*
  5. *If $a \in \mathbb{F}^\alpha C_n y$, then $\widehat{a} \in \mathbb{F}^\alpha C_n y$.*

*Proof.* The assertions can be checked straightforwardly.
$\square$

**Definition 3.2.**  1. For a 2-cocycle $\alpha$ of $D_{2n}$ we define the reversible subspace of $\mathbb{F}^\alpha C_n y$ as the vector subspace

$$\Gamma_\alpha = \{a = \sum_{i=0}^{n-1} a_i \overline{x^i y} \in \mathbb{F}^\alpha C_n y \mid a_i = a_{n-i} \text{ for } i = 1, \ldots, n-1\}.$$

  2. Given $a = \sum_{i=0}^{n-1} a_i \overline{x^i y} \in \mathbb{F}^\alpha C_n y$ we define $\Phi(a) = \sum_{i=0}^{n-1} a_i \overline{x^i} \in \mathbb{F}^\alpha C_n$.

Note that the map $\Phi : \mathbb{F}^\alpha C_n y \longrightarrow \mathbb{F}^\alpha C_n$ is an $\mathbb{F}$-linear isomorphism.

**Lemma 3.3.** *Let $\alpha$ be a 2-cocycle of $D_{2n}$, then we have*
  1. *If*
$$\alpha(x^i, x^{j-i}) = \alpha(x^{j-i}, x^i) \tag{1}$$
  *for all $i, j \in \{0, \ldots, n-1\}$, then $ab = ba$ for $a, b \in \mathbb{F}^\alpha C_n$.*

2. If
$$\alpha(x^{i-j}y, x^{i-j}y)\alpha(x^i y, x^{i-j}y) = \alpha(x^{n-i}y, x^{n-i}y)\alpha(x^{j-i}y, x^{n-i}y) \quad (2)$$
for all $i,j \in \{0,\ldots,n-1\}$, then $\widehat{ab} = \widehat{ba}$ for $a,b \in \Gamma_\alpha$.

*Proof.* 1) Let $a,b \in \mathbb{F}^\alpha C_n$, then $ab = \sum_{i=0}^{n-1} c_j \overline{x^j}$, where
$$c_j = \sum_{i=0}^{n-1} a_i b_{[j-i]_n} \alpha(x^i, x^{j-i})$$
and $ba = \sum_{j=0}^{n-1} d_j \overline{x^j}$, where
$$d_j = \sum_{i=0}^{n-1} b_i a_{[j-i]_n} \alpha(x^i, x^{j-i}).$$

Let us fix $j$ and consider the term $i \in \{0,\ldots,n-1\}$ in the summation $c_j$, viz.
$$c_j^i = a_i b_{[j-i]_n} \alpha(x^i, x^{j-i}),$$
and the term $[j-i]_n$ in the summation $d_j$, viz.
$$d_j^{[j-i]_n} = b_{[j-i]_n} a_i \alpha(x^{j-i}, x^i).$$

Therefore, $c_j^i = d_j^{[j-i]_n}$ for all $i,j \in \{0,\ldots,n-1\}$ and the assertion follows.

2) Since $a \in \Gamma_\alpha$, then
$$a = a_0 \overline{y} + a_1 \overline{xy} + a_2 \overline{x^2 y} + \cdots + a_2 \overline{x^{n-2}y} + a_1 \overline{x^{n-1}y}.$$

By Item 5 of Lemma 3.1, then
$$\widehat{a} = a_0 \alpha(y,y)\overline{y} + a_1 \alpha(xy,xy)\overline{xy} + \cdots + a_2 \alpha(x^{n-2}y, x^{n-2}y)\overline{x^{n-2}y} + a_1 \alpha(x^{n-1}y, x^{n-1}y)\overline{x^{n-1}y}.$$

Similarly, since $b \in \Gamma_\alpha$, then
$$b = b_0 \overline{y} + b_1 \overline{xy} + b_2 \overline{x^2 y} + \cdots + b_2 \overline{x^{n-2}y} + b_1 \overline{x^{n-1}y},$$
and
$$\widehat{b} = b_0 \alpha(y,y)\overline{y} + b_1 \alpha(xy,xy)\overline{xy} + \cdots + b_2 \alpha(x^{n-2}y, x^{n-2}y)\overline{x^{n-2}y} + b_1 \alpha(x^{n-1}y, x^{n-1}y)\overline{x^{n-1}y}.$$

Therefore $\widehat{a}\widehat{b} = \sum_{j=0}^{n-1} c_j \overline{x^j}$, where
$$c_j = \sum_{i=0}^{n-1} a_i b_{[i-j]_n} \alpha(x^{i-j}y, x^{i-j}y) \alpha(x^i y, x^{i-j}y),$$
and $\widehat{b}\widehat{a} = \sum_{j=0}^{n-1} d_j \overline{x^j}$, where
$$d_j = \sum_{i=0}^{n-1} b_i a_{[i-j]_n} \alpha(x^{i-j}y, x^{i-j}y) \alpha(x^i y, x^{i-j}y).$$

Let us fix $j$ and consider the term $i \in \{0,1,\ldots,n-1\}$ in the summation $c_j$, viz.
$$c_j^i = a_i b_{[i-j]_n} \alpha(x^{i-j}y, x^{i-j}y) \alpha(x^i y, x^{i-j}y),$$
and the term $[j-i]_n$ in the summation $d_j$, viz.
$$d_j^{[j-i]_n} = b_{[j-i]_n} a_{[-i]_n} \alpha(x^{n-i}y, x^{n-i}y) \alpha(x^{j-i}y, x^{n-i}y).$$

Since $a, b \in \Gamma_\alpha$, then $a_i = a_{[-i]_n}$ and $b_{[j-i]_n} = b_{[i-j]_n}$, then $c_j^i = d_j^{[j-i]_n}$ for all $i,j \in \{0,\ldots,n-1\}$ and the assertion follows.

□

In the following lemma we define a particular 2-cocycle of $D_{2n}$, which we will use to construct a twisted dihedral group algebra.

**Lemma 3.4.** *Let $\lambda$ be an element in $\mathbb{F}^* = \mathbb{F} \setminus \{0\}$. The map $\alpha_\lambda : D_{2n} \times D_{2n} \longrightarrow \mathbb{F}^*$ defined by $\alpha_\lambda(g, h) = \lambda$ for $g = x^i y$, $h = x^j y$ with $i, j \in \{0, \ldots, n-1\}$ and $\alpha_\lambda(g, h) = 1$ otherwise is a 2-cocycle.*

*Proof.* By definition $\alpha_\lambda(1, 1) = 1$. Therefore $\alpha_\lambda$ is a 2-cocycle if $\alpha_\lambda(g, h)\alpha_\lambda(gh, k) = \alpha_\lambda(g, hk)\alpha_\lambda(h, k)$ for all $g, h, k \in D_{2n}$. Let us first assume $g = x^i$, $h = x^{j_1} y^{k_1}$ and $k = x^{j_2} y^{k_2}$ with $i, j_1, j_2 \in \{0, \ldots, n-1\}$ and $k_1, k_2 \in \{0, 1\}$, then an straightforward calculation shows that $\alpha_\lambda(g, h)\alpha_\lambda(gh, k) = \alpha_\lambda(g, hk)\alpha_\lambda(h, k)$ holds. On the other hand, if $g = x^i y$, then a straightforward calculation also shows that $\alpha_\lambda(g, h)\alpha_\lambda(gh, k) = \alpha_\lambda(g, hk)\alpha_\lambda(h, k)$ holds. □

**Lemma 3.5.** $\mathbb{F}D_{2n}$ *and* $\mathbb{F}^{\alpha_\lambda} D_{2n}$ *are isomorphic if and only if $\lambda$ is a square in $\mathbb{F}$.*

*Proof.* Suppose $\lambda$ is a square in $\mathbb{F}$, then there exists $t \in \mathbb{F}^*$ with $t^2 = \lambda$. By setting $\beta(x^i) = 1$ and $\beta(x^i y) = t^{-1}$ for all $i \in \{0, 1, \ldots, n-1\}$, we clearly have $\alpha_\lambda(g, h) = \beta(g)^{-1}\beta(h)^{-1}\beta(gh)$ for all $g, h \in D_{2n}$.

Now suppose that $\lambda$ is not a square in $\mathbb{F}$ and that there exists $\beta : G \longrightarrow \mathbb{F}^*$ with $\beta(1) = 1$ such that $\alpha_\lambda(g, h) = \beta(g)^{-1}\beta(h)^{-1}\beta(gh)$. By definition $\alpha_\lambda(y, y) = \lambda$. Moreover $\alpha_\lambda(y, y) = \beta(y)^{-1}\beta(y)^{-1} = [\beta(y)^{-1}]^2$, a contradiction. □

**Remark 2.** Since $(\lambda^{2^{m-1}})^2 = \lambda$ for all $\lambda \in \mathbb{F}_{2^m}$, then $\mathbb{F}_{2^m} D_{2n}$ and $\mathbb{F}_{2^m}^{\alpha_\lambda} D_{2n}$ are isomorphic for any $\lambda$ and $m \in \mathbb{N}$. For $p > 2$, by generalized Euler's criterion we have $\lambda$ is a square in $\mathbb{F}_{p^m}$ if and only $\lambda^{(p^m-1)/2} = 1$. Therefore $\mathbb{F}_{p^m} D_{2n}$ and $\mathbb{F}_{p^m}^{\alpha_\lambda} D_{2n}$ are isomorphic if and only if $\lambda^{(p^m-1)/2} = 1$.

**Lemma 3.6.** *If $\lambda_1, \lambda_2$ are not squares in $\mathbb{F}$, then $\mathbb{F}^{\alpha_{\lambda_1}} D_{2n}$ and $\mathbb{F}^{\alpha_{\lambda_2}} D_{2n}$ are isomorphic.*

*Proof.* By Remark 2 we may assume the characteristic of $\mathbb{F}$ is not 2. Let $\xi$ be a primitive element of $\mathbb{F}$ and since $\lambda_1, \lambda_2$ are not squares, then $\lambda_1 = \xi^{k_1}$ and $\lambda_2 = \xi^{k_2}$ with $k_1$ and $k_2$ being odd. Therefore $\lambda_1 = \xi^{k_1} = \lambda_2 \xi^{k_3}$, where $k_3$ is even, i.e. $\xi^{k_3}$ is a square. By setting $\beta(x^i) = 1$ and $\beta(x^i y) = \xi^{k_3/2}$ for all $i \in \{0, 1, \ldots, n-1\}$, we clearly have $\alpha_{\lambda_1}(g, h) = \alpha_{\lambda_2}(g, h)\beta(g)\beta(h)\beta(gh)^{-1}$ for all $g, h \in D_{2n}$. □

**Lemma 3.7.** *The reversible subset $\Gamma_{\alpha_\lambda}$ of $\mathbb{F}^{\alpha_\lambda} C_n y$ is self-adjoint, viz. $\Gamma_{\alpha_\lambda} = \widehat{\Gamma_{\alpha_\lambda}}$.*

*Proof.* Let $\widehat{a} = \sum_{i=0}^{n-1} a_i \alpha_\lambda(x^i y, x^i y) \overline{x^i y} \in \widehat{\Gamma_{\alpha_\lambda}}$. Since
$$a_i \alpha_\lambda(x^i y, x^i y) = a_{n-i} \alpha_\lambda(x^{n-i} y, x^{n-i} y),$$
then $\widehat{a} \in \Gamma_{\alpha_\lambda}$. On the other hand, if $a = \sum_{i=0}^{n-1} a_i \overline{x^i y} \in \Gamma_{\alpha_\lambda}$ we have that
$$a' = \sum_{i=0}^{n-1} a_i \alpha_\lambda(x^i y, x^i y)^{-1} \overline{x^i y} \in \widehat{\Gamma}$$
and $a = \widehat{a'}$. □

**Lemma 3.8.** *Let $a, b$ be elements in $\mathbb{F}^{\alpha_\lambda} C_n y$, then the following assertions hold.*
  1. *If $a \in \Gamma_{\alpha_\lambda}$, then $a = \Phi(a)\overline{y} = \overline{y}\Phi(a)$.*
  2. *$a \in \Gamma_{\alpha_\lambda}$ if and only if $\Phi(a) = \widehat{\Phi(a)}$.*

3. If $a \in \Gamma_{\alpha_\lambda}$, then $\widehat{\Phi(a)\overline{y}} = \widehat{\overline{y}}\Phi(a)$.
4. $\Phi(a)\Phi(b) = \Phi(b)\Phi(a)$.
5. If $a, b \in \Gamma_{\alpha_\lambda}$, then $a\widehat{b} = b\widehat{a}$.

*Proof.* 1.
$$a = \sum_{i=0}^{n-1} a_i \overline{x^i y} = \sum_{i=0}^{n-1} a_i \alpha_\lambda(x^i, y) \overline{x^i y} = \Phi(a)\overline{y}.$$

Also
$$\overline{y}\Phi(a) = \sum_{i=0}^{n-1} a_i \overline{y x^i} = \sum_{i=0}^{n-1} a_i \overline{x^{n-i} y} = \sum_{i=0}^{n-1} a_{[n-i]_n} \overline{x^{n-i} y} = \sum_{i=0}^{n-1} a_i \overline{x^i y} = \sum_{i=0}^{n-1} a_i \overline{x^i} \overline{y} = \Phi(a)\overline{y}.$$

2. $a \in \Gamma_{\alpha_\lambda} \Leftrightarrow a_i = a_{[n-i]_n} \Leftrightarrow a_i = a_{[n-i]_n} \alpha_\lambda(x^{n-i}, x^i) \Leftrightarrow \Phi(a) = \widehat{\Phi(a)}$.
3. Given $a \in \Gamma_{\alpha_\lambda}$ then we get

$$\widehat{\Phi(a)\overline{y}} = \sum_{i=0}^{n-1} a_i \alpha_\lambda(x^i, y) \alpha_\lambda(x^i y, x^i y) \overline{x^i y} = \lambda a$$

and

$$\widehat{\overline{y}}\Phi(a) = \alpha_\lambda(y, y)\overline{y} \sum_{i=0}^{n-1} a_i \overline{x^i} = \lambda \sum_{i=0}^{n-1} a_i \alpha_\lambda(y, x^i) \overline{x^{n-i} y} = \lambda \sum_{i=0}^{n-1} a_i \alpha_\lambda(y, x^i) \overline{x^i y} = \lambda a.$$

4. Since $\Phi(a), \Phi(b) \in \mathbb{F}^{\alpha_\lambda} C_n$ and $\alpha_\lambda$ satisfies Equation (1) the assertion follows.
5. $a\widehat{b} = \Phi(a)\overline{y}\widehat{\Phi(b)\overline{y}} = \Phi(a)\overline{y}\widehat{\overline{y}}\Phi(b) = \lambda^2 \Phi(a)\Phi(b) = \lambda^2 \Phi(b)\Phi(a) = \Phi(b)\overline{y}\widehat{\overline{y}}\Phi(a) = b\widehat{a}$.

□

4. **A key exchange protocol.** In this section, we present a key exchange protocol based on a twisted dihedral group algebra. We remark that there are other works that have considered two-sided semigroup actions or matrices over group rings to do key exchange in the manner or style of Diffie-Hellman [14, 17, 16]. However our work uses two-sided multiplications over a twisted dihedral group algebra to do a key exchange protocol.

4.1. **Cryptographic Construction.** We start off by describing how to set up our protocol's public parameters.

1. Choose $m \in \mathbb{N}$ and a number prime $p$ with $p|2n$ and set $q = p^m$.
2. Choose a 2-cocycle $\alpha$ such that $\mathbb{F}_q^\alpha D_{2n}$ and $\mathbb{F}_q D_{2n}$ are not isomorphic. This ensures all protocol's operations are carried out in a twisted dihedral group algebra, namely $\mathbb{F}_q^\alpha D_{2n}$. Additionally, $\alpha$ must satisfy Equation (1) and Equation (2) to guarantee $\mathtt{a}_1 \mathtt{a}_2 = \mathtt{a}_2 \mathtt{a}_1$ if $\mathtt{a}_1, \mathtt{a}_2 \in \mathbb{F}_q^\alpha C_n$ and $\gamma_1 \widehat{\gamma_2} = \gamma_2 \widehat{\gamma_1}$ if $\gamma_1, \gamma_2 \in \Gamma_\alpha$. In particular, we choose the 2-cocycle $\alpha_\lambda$ from Lemma 3.4 for a non-square $\lambda$ in $\mathbb{F}_q$ with $p > 2$. We can find $\lambda$ efficiently by selecting a random element $\lambda \in \mathbb{F}_q^*$ and verifying if $\lambda^{(q-1)/2} = -1$.
3. Choose a random non-zero element $\mathtt{h}_1 \in \mathbb{F}_q^{\alpha_\lambda} C_n$ and a random non-zero element $\mathtt{h}_2 \in \mathbb{F}_q^{\alpha_\lambda} C_n y$. Set $\mathtt{h} = \mathtt{h}_1 + \mathtt{h}_2$ and make $\mathtt{h}$ public.

We use the notation introduced in [3]. Let $P_i$ and $P_j$ be two parties and $s$ be an identifier for a session. The key exchange protocol between $P_i$ and $P_j$ runs as shown by Protocol 1.

**Protocol 1** Our Key Exchange Protocol

1: The initiator $P_i$, on input $(P_i, P_j, s)$, chooses a secret pair $(\mathtt{a}_i, \gamma_i) \xleftarrow{R} \mathbb{F}_q^{\alpha_\lambda} C_n \times \Gamma_{\alpha_\lambda}$, and sends $(P_i, s, \mathtt{pk}_i = \mathtt{a}_i \mathtt{h} \gamma_i)$ to $P_j$.
2: Upon receipt of $(P_i, s, \mathtt{pk}_i)$ the responder, $P_j$, chooses a secret pair $(\mathtt{a}_j, \gamma_j) \xleftarrow{R} \mathbb{F}_q^{\alpha_\lambda} C_n \times \Gamma_{\alpha_\lambda}$ and sends $(P_j, s, \mathtt{pk}_j = \mathtt{a}_j \mathtt{h} \gamma_j)$ to $P_i$, computes $\mathtt{k}_j = \mathtt{a}_j \mathtt{pk}_i \widehat{\gamma_j}$, erases $(\mathtt{a}_j, \gamma_j)$ and outputs the key $\mathtt{k}_j$ under the session-id $s$.
3: Upon receipt of $(P_j, s, \mathtt{pk}_j)$ the party, $P_i$, computes $\mathtt{k}_i = \mathtt{a}_i \mathtt{pk}_j \widehat{\gamma_i}$, erases $(\mathtt{a}_i, \gamma_i)$ and outputs the key $\mathtt{k}_i$ under the session-id $s$.

Note that if both $P_i$ and $P_j$ are uncorrupted during the exchange of the key and both complete the protocol for session-id $s$, then they both establish the same key. Indeed, since the choice of $\alpha_\lambda$, we have $\mathtt{a}_i \mathtt{a}_j = \mathtt{a}_j \mathtt{a}_i$ and $\gamma_i \widehat{\gamma_j} = \gamma_j \widehat{\gamma_i}$, then

$$\mathtt{k}_i = \mathtt{a}_i \mathtt{pk}_j \widehat{\gamma_i} = \mathtt{a}_i \mathtt{a}_j \mathtt{h} \gamma_j \widehat{\gamma_i} = \mathtt{a}_j \mathtt{a}_i \mathtt{h} \gamma_i \widehat{\gamma_j} = \mathtt{a}_j \mathtt{pk}_i \widehat{\gamma_j} = \mathtt{k}_j$$

In Section 4.5, we analyse further this key exchange protocol in the authenticated-links adversarial model of Canetti and Krawczyk [3].

4.2. **Comparison with other approaches.** The authors of [10] introduce a key exchange protocol based on a twisted dihedral group algebra. They first introduce a map $\beta_\lambda : D_{2n} \times D_{2n} \to \mathbb{F}^*$ defined as $\beta_\lambda(x^i, x^j y^k) = 1$ and $\beta_\lambda(x^i y, x^j y^k) = \lambda^j$ for $i, j \in \{0, 1, \ldots, n-1\}$, $k \in \{0, 1\}$ and $\lambda \in \mathbb{F}^*$. They claim the map $\beta_\lambda$ to be a 2-cocycle by choosing $\lambda$ as a primitive root over $\mathbb{F}^*$. Also, they construct the twisted group algebra $\mathbb{F}^{\beta_\lambda} D_{2n}$ and define the operator $^* : \mathbb{F}^{\beta_\lambda} D_{2n} \longrightarrow \mathbb{F}^{\beta_\lambda} D_{2n}$ defined as

$$a^* = \sum_{i \in \{0, 1, \ldots, n-1\}, k \in \{0, 1\}} a_{i,k} \lambda^{-i} \overline{x^i y^k}.$$

They prove that $ab = ba$ if $a, b \in \mathbb{F}^{\beta_\lambda} C_n$ and $ab^* = ba^*$ if $a, b \in \Gamma_{\beta_\lambda}$. Equipped with the asterisk operator, then they construct the following protocol. The participant $P_i$ ($i \in \{0, 1\}$) randomly selects her/his secret key as $(\mathtt{a}_i, \gamma_i)$, where $\mathtt{a}_i \in \mathbb{F}_q^{\beta_\lambda} C_n$ and $\gamma_i \in \Gamma_{\beta_\lambda}$, then computes the corresponding public key as $\mathtt{pk}_i = \mathtt{a}_i h \gamma_i$ and sends it to her/his counterpart. Upon receiving the public key $\mathtt{pk}_{1-i}$ from her/his peer, the participant $P_i$ finally computes the shared key by calculating $\mathtt{k} = \mathtt{a}_i \mathtt{pk}_{1-i} \gamma_i^*$.

We first remark that for the map $\beta_\lambda$ to be a 2-cocycle, the order of $\lambda$ in $\mathbb{F}^*$, denoted as $\mathrm{ord}(\lambda)$, must satisfy some constraints.

**Lemma 4.1.** *The map $\beta_\lambda$ is a 2-cocycle if and only if $\mathrm{ord}(\lambda)$ divides $n$.*

*Proof.* Note that $\beta_\lambda(x^i y^k, 1) = \beta_\lambda(1, x^i y^k) = 1$, for all $x^i, y^k \in D_{2m}$, with $i \in \{0, 1, \ldots, n-1\}$ and $k \in \{0, 1\}$ by construction. Therefore, $\beta_\lambda$ is 2-cocycle if $\beta_\lambda(g_1, g_2) \beta_\lambda(g_1 g_2, g_3) = \beta_\lambda(g_1, g_2 g_3) \beta_\lambda(g_2, g_3)$ for all $g_1, g_2, g_3 \in D_{2m}$

Suppose that $\mathrm{ord}(\lambda)$ does not divide $n$. Therefore, $n = q \cdot \mathrm{ord}(\lambda) + r$, with $0 < r < \mathrm{ord}(\lambda)$. Consider $g_1 = y, g_2 = x, g_3 = x^{n-1}$. On the one hand, $\beta_\lambda(g_1, g_2) \beta_\lambda(g_1 g_2, g_3) = \beta_\lambda(y, x) \beta_\lambda(yx, x^{n-1}) = \beta_\lambda(y, x) \beta_\lambda(x^{n-1} y, x^{n-1}) = \lambda \cdot \lambda^{n-1} = \lambda^{q \cdot \mathrm{ord}(\lambda) + r} = \lambda^r \neq 1$. On the other hand,

$$\beta_\lambda(g_1, g_2 g_3) \beta_\lambda(g_2, g_3) = \beta_\lambda(y, x^n) \beta_\lambda(x, x^{n-1}) = \beta_\lambda(y, 1) \beta_\lambda(x, x^{n-1}) = 1 \cdot 1 = 1.$$

Therefore, $\beta_\lambda(y, x) \beta_\lambda(yx, x^{n-1}) \neq \beta_\lambda(y, x^n) \beta_\lambda(x, x^{n-1})$ and so $\beta_\lambda$ is not a 2-cocycle.

Now suppose that $\mathrm{ord}(\lambda)$ divides $n$. We proceed by analysing two cases:

- Case 1: Suppose $g_1 = x^i, g_2 = x^j y^{k_1}, g_3 = x^l y^{k_2}$, with $i, j, l \in \{0, 1, \ldots, n-1\}$ and $k_1, k_2 \in \{0, 1\}$.

    Firstly, consider

$$\beta_\lambda(g_1, g_2)\beta_\lambda(g_1 g_2, g_3) = \beta_\lambda(x^i, x^j y^{k_1})\beta_\lambda(x^{i+j} y^{k_1}, x^l y^{k_2}) = \beta_\lambda(x^{i+j} y^{k_1}, x^l y^{k_2}).$$

   Thus, it results in $\lambda^l$ if $k_1 = 1$ or 1 otherwise.
   Secondly, consider

$$\beta_\lambda(g_1, g_2 g_3)\beta_\lambda(g_2, g_3) = \beta_\lambda(x^i, x^j y^{k_1} x^l y^{k_2})\beta_\lambda(x^j y^{k_1}, x^l y^{k_2}) = \beta_\lambda(x^i y^{k_1}, x^l y^{k_2}).$$

   Thus, it results in $\lambda^l$ if $k_1 = 1$ or 1 otherwise. Therefore,

$$\beta_\lambda(g_1, g_2)\beta_\lambda(g_1 g_2, g_3) = \beta_\lambda(g_1, g_2 g_3)\beta_\lambda(g_2, g_3).$$

- Case 2: Suppose $g_1 = x^i, g_2 = x^j y^{k_1}, g_3 = x^l y^{k_2}$, with $i, j, l \in \{0, 1, \ldots, n-1\}$ and $k_1, k_2 \in \{0, 1\}$.
   Firstly, consider

$$\beta_\lambda(g_1, g_2)\beta_\lambda(g_1 g_2, g_3) = \beta_\lambda(x^i y, x^j y^{k_1})\beta_\lambda(x^i y x^j y^{k_1}, x^l y^{k_2}) = \lambda^j \beta_\lambda(x^i x^{n-j} y^{k_1 + 1}, x^l y^{k_2}).$$

   Thus, it results in $\lambda^{l+j}$ if $k_1 = 0$ or $\lambda^j$ otherwise.
   Secondly, consider $\beta_\lambda(g_1, g_2 g_3)\beta_\lambda(g_2, g_3)$, thus

$$\beta_\lambda(g_1, g_2 g_3)\beta_\lambda(g_2, g_3) = \beta_\lambda(x^i y, x^j y^{k_1} x^l y^{k_2})\beta_\lambda(x^j y^{k_1}, x^l y^{k_2}).$$

   Therefore, it results in $\lambda^{j+l}$ if $k_1 = 0$ or $\lambda^{j+n}$ otherwise. Since $\text{ord}(\lambda)$ divides $n$, then $\lambda^{j+n} = \lambda^j$. Therefore,

$$\beta_\lambda(g_1, g_2)\beta_\lambda(g_1 g_2, g_3) = \beta_\lambda(g_1, g_2 g_3)\beta_\lambda(g_2, g_3).$$

We conclude that $\beta_\lambda$ is a 2-cocycle. □

Consider the prime factorization of $\gcd(|\mathbb{F}| - 1, n) = p_1^{r_1} p_2^{r_2} \cdots p_k^{r_k}$, where $p_i$ are prime numbers and $r_i$ are positive integers. Thus, there are $\prod_{i=1}^{k}(r_i + 1)$ divisors for both $|\mathbb{F}| - 1$ and $n$. By finding a field element $\lambda \in \mathbb{F}^*$ such that $\text{ord}(\lambda)$ divides $\gcd(|\mathbb{F}| - 1, n)$, then the 2-cocycle $\beta_\lambda$ and the corresponding twisted group algebra $\mathbb{F}^{\beta_\lambda} D_{2n}$ can be constructed. Note that if $\gcd(|\mathbb{F}| - 1, n) = 1$, then any suitable candidate for $\lambda$ must have its order dividing 1, then the only possible candidate for $\lambda$ is 1. As a result, $\mathbb{F} D_{2n}$ and $\mathbb{F}^{\beta_1} D_{2n}$ are isomorphic. This is simply a special case of a more general result.

**Lemma 4.2.** *Let $\lambda_1, \lambda_2 \in \mathbb{F}^*$ be two field elements. If $\lambda_1, \lambda_2$ are squares and $\text{ord}(\lambda_1)$ divides $n$, then $\mathbb{F}^{\beta_{\lambda_1}} D_{2n}$ and $\mathbb{F}^{\alpha_{\lambda_2}} D_{2n}$ are isomorphic.*

*Proof.* Since $\lambda_1$ and $\lambda_2$ are squares in $\mathbb{F}$, then there exists $\xi_1, \xi_2 \in \mathbb{F}$ such that $\lambda_1 = \xi_1^2$ and $\lambda_2 = \xi_2^2$. Also, $\lambda_1^i = \xi_1^{2i}$ for all $0 \le i < \text{ord}(\lambda_1)$. By setting $\theta(x^i) = 1/\xi_1^i$ and $\theta(x^i y) = \xi_2/\xi_1^i$ for all $i \in \{0, 1, \ldots, n-1\}$, we clearly have $\alpha_{\lambda_2}(g, h) = \beta_{\lambda_1}(g, h)\theta(g)\theta(h)\theta(gh)^{-1}$ for all $g, h \in D_{2n}$. □

From previous lemma, it follows that if the order of $\lambda_1$ is odd and divides $\gcd(|\mathbb{F}| - 1, n)$, then $\mathbb{F}^{\beta_{\lambda_1}} D_{2n}$ and $\mathbb{F} D_{2n}$ are isomorphic. Indeed, suppose $\text{ord}(\lambda_1)$ is odd and divides $\gcd(|\mathbb{F}| - 1, n)$, then $\lambda_1$ is a square, since $\xi_1^2 = \lambda_1$ with $\xi_1 = \lambda_1^{(\text{ord}(\lambda_1)+1)/2}$, and hence $\mathbb{F}^{\beta_{\lambda_1}} D_{2n}$ and $\mathbb{F}^{\alpha_{\lambda_2}} D_{2n}$ are isomorphic for some square $\lambda_2 \in \mathbb{F}^*$ (in particular, $\lambda_2$ may be set to 1). Furthermore, if $n$ is odd or $|\mathbb{F}| - 1$ is odd, then $\text{ord}(\lambda_1)$ is odd for any $\lambda_1$ with order dividing $\gcd(|\mathbb{F}| - 1, n)$. Therefore, $\mathbb{F}^{\beta_{\lambda_1}} D_{2n}$ and $\mathbb{F} D_{2n}$ are isomorphic for any $\lambda_1$ with order dividing $\gcd(|\mathbb{F}| - 1, n)$.

In the following, we will investigate the structure of $\mathbb{F}^{\beta_\lambda} D_{2n}$ when $\lambda$ is a non-square.

**Lemma 4.3.** *Let $\lambda_1, \lambda_2 \in \mathbb{F}^*$ be two field elements. If $\lambda_1, \lambda_2$ are non-squares such that both $\mathrm{ord}(\lambda_1)$ and $\mathrm{ord}(\lambda_2)$ divide $n$, then the 2-cocycles $\beta_{\lambda_1}$ and $\beta_{\lambda_2}$ are equivalent. In particular, $\mathbb{F}^{\beta_{\lambda_1}} D_{2n}$ and $\mathbb{F}^{\beta_{\lambda_2}} D_{2n}$ are isomorphic.*

*Proof.* Let $\xi$ be a primitive element of $\mathbb{F}$ and since $\lambda_1, \lambda_2$ are not squares, then $\lambda_1 = \xi^{k_1}$ and $\lambda_2 = \xi^{k_2}$ with $k_1$ and $k_2$ being odd. Therefore $\lambda_1 = \xi^{k_1} = \lambda_2 \xi^{k_3}$, where $k_3$ is even, i.e. $\xi^{k_3}$ is a square.

By setting $\theta(x^i y^k) = \xi^{i \cdot k_3/2}$ for all $i \in \{0, 1, \ldots, n-1\}, k \in \{0, 1\}$, we clearly have $\beta_{\lambda_1}(g, h) = \beta_{\lambda_2}(g, h)\theta(g)\theta(h)\theta(gh)^{-1}$ for all $g, h \in D_{2n}$. $\square$

We now prove a result stating that $\alpha_{\lambda_1}$ for a non-square $\lambda_1 \in \mathbb{F}^*$ and $\beta_{\lambda_2}$ for any suitable $\lambda_2 \in \mathbb{F}^*$ are not equivalent. This allows us to argue each protocol operates on a structurally different twisted group algebra.

**Lemma 4.4.** *Let $\lambda_1, \lambda_2 \in \mathbb{F}^*$ be two field elements. If $\lambda_1$ is a non-square in $\mathbb{F}$ and $\mathrm{ord}(\lambda_2)$ divides $n$, then the 2-cocycles $\alpha_{\lambda_1}$ and $\beta_{\lambda_2}$ are not equivalent. In particular, $\mathbb{F}^{\alpha_{\lambda_1}} D_{2n}$ and $\mathbb{F}^{\beta_{\lambda_2}} D_{2n}$ are not isomorphic.*

*Proof.* Suppose that there exists $\theta : G \longrightarrow \mathbb{F}^*$ with $\theta(1) = 1$ such that
$$\alpha_{\lambda_1}(g, h) = \beta_{\lambda_2}(g, h)\theta(g)\theta(h)\theta(gh)^{-1}.$$
By definition $\alpha_{\lambda_1}(y, y) = \lambda_1$ and $\beta_{\lambda_2}(y, y) = \lambda_2^0 = 1$. Moreover
$$\alpha_{\lambda_1}(y, y) = \lambda_1 = \beta_{\lambda_2}(y, y)\theta(y)\theta(y)\theta(1)^{-1} = [\theta(y)]^2,$$
which is a contradiction. $\square$

4.3. **Intractability assumptions.** We now present attack games [22] for algebraic problems related to the security of our key exchange protocol.

Let $\mathbb{F}_q$ be a finite field with $q = p^m$ and $p > 2$, $D_{2n}$ be the dihedral group of order $2n$, $\alpha_\lambda$ be the 2-cocycle from Lemma 3.4 for a non-square $\lambda \in \mathbb{F}_q$, and $\mathtt{h} = \mathtt{h}_1 + \mathtt{h}_2$ be a public element in $\mathbb{F}_q^{\alpha_\lambda} D_{2n}$, where $\mathtt{h}_1$ is a random non-zero element from $\mathbb{F}_q^{\alpha_\lambda} C_n$ and $\mathtt{h}_2$ is a random non-zero element from $\mathbb{F}_q^{\alpha_\lambda} C_n y$.

**Game 1** (Dihedral Product Decomposition). *For a given adversary $\mathcal{A}$, we define the following attack game:*

- *The challenger computes*
  1: $(\mathtt{a}, \gamma) \xleftarrow{R} \mathbb{F}_q^{\alpha_\lambda} C_n \times \Gamma_{\alpha_\lambda}$;
  2: $\mathtt{pk} \leftarrow \mathtt{a}\mathtt{h}\gamma$;
  
  *and gives the value of $\mathtt{pk}$ to the adversary.*
- *The adversary outputs $(\widetilde{\mathtt{a}}, \widetilde{\gamma}) \in \mathbb{F}_q^{\alpha_\lambda} C_n \times \Gamma_{\alpha_\lambda}$.*

We define $\mathcal{A}$'s advantage in solving the Dihedral Product Decomposition Problem for $\mathbb{F}_q^{\alpha_\lambda} D_{2n}$, denoted $\mathrm{DPDadv}[\mathcal{A}, \mathbb{F}_q^{\alpha_\lambda} D_{2n}]$, as the probability that $\widetilde{\mathtt{a}}\mathtt{h}\widetilde{\gamma} = \mathtt{a}\mathtt{h}\gamma$.

**Definition 4.5** (Dihedral Product Decomposition Assumption). We say that the Dihedral Product Decomposition (DPD) assumption holds for $\mathbb{F}_q^{\alpha_\lambda} D_{2n}$ if for all efficient adversaries $\mathcal{A}$ the quantity $\mathrm{DPDadv}[\mathcal{A}, \mathbb{F}_q^{\alpha_\lambda} D_{2n}]$ is negligible.

Let us assume that the DPD assumption does not hold for $\mathbb{F}_q^{\alpha_\lambda} D_{2n}$. Therefore, if two parties $P_1$ and $P_2$ are uncorrupted during the exchange of the key and both complete the protocol, then an adversary, which passively captures $\text{pk}_1 = a_1 h \gamma_1$ or $\text{pk}_2 = a_2 h \gamma_2$ or both, can find $(\widetilde{a}, \widetilde{\gamma}) \in \mathbb{F}_q^{\alpha_\lambda} C_n \times \Gamma_{\alpha_\lambda}$ with $\widetilde{a} h \widetilde{\gamma} = a_1 h \gamma_1 = \text{pk}_1$ with non-negligible probability, since the DPD assumption does not hold for $\mathbb{F}_q^{\alpha_\lambda} D_{2n}$, and therefore can easily compute $\widetilde{a} \text{pk}_2 \widehat{\widetilde{\gamma}} = a_2 \widetilde{a} h \widetilde{\gamma} \widehat{\gamma}_2 = a_2 \text{pk}_A \widehat{\gamma}_2 = \text{k}$.

We now present other algebraic problems related to the DPD problem.

**Game 2** (Computational Dihedral Product). *For a given adversary $\mathcal{A}$, we define the following attack game:*

- *The challenger computes*
  1: $(a_1, \gamma_1) \xleftarrow{R} \mathbb{F}_q^{\alpha_\lambda} C_n \times \Gamma_{\alpha_\lambda}$;
  2: $(a_2, \gamma_2) \xleftarrow{R} \mathbb{F}_q^{\alpha_\lambda} C_n \times \Gamma_{\alpha_\lambda}$;
  3: $pk_1 \leftarrow a_1 h \gamma_1$;
  4: $pk_2 \leftarrow a_2 h \gamma_2$;
  5: $k \leftarrow a_2 pk_1 \widehat{\gamma}_2$;
  *and gives the values of $pk_1$ and $pk_2$ to the adversary.*
- *The adversary outputs some $\widetilde{k} \in \mathbb{F}_q^{\alpha_\lambda} D_{2n}$*

*We define $\mathcal{A}$'s advantage in solving the Computational Dihedral Product (CDP) Problem for $\mathbb{F}_q^{\alpha_\lambda} D_{2n}$, denoted CDPadv$[\mathcal{A}, \mathbb{F}_q^{\alpha_\lambda} D_{2n}]$, as the probability that $\widetilde{k} = k$.*

**Definition 4.6** (Computational Dihedral Product Assumption). We say that the Computational Dihedral Product (CDP) assumption holds for $\mathbb{F}_q^{\alpha_\lambda} D_{2n}$ if for all efficient adversaries $\mathcal{A}$ the quantity CDPadv$[\mathcal{A}, \mathbb{F}_q^{\alpha_\lambda} D_{2n}]$ is negligible.

Note that if an adversary wins the DPD game with non-negligible probability for $\mathbb{F}_q^{\alpha_\lambda} D_{2n}$, then such adversary also wins the CDP game with non-negligible probability for $\mathbb{F}_q^{\alpha_\lambda} D_{2n}$.

We now present a stronger computational problem, the decisional version of the CDP problem.

**Game 3** (Decisional Dihedral Product). *For a given adversary $\mathcal{A}$, we define two experiments:*

**Experiment $b$**

- *The challenger computes*
  1: $(a_1, \gamma_1) \xleftarrow{R} \mathbb{F}_q^{\alpha_\lambda} C_n \times \Gamma_{\alpha_\lambda}$;
  2: $(a_2, \gamma_2) \xleftarrow{R} \mathbb{F}_q^{\alpha_\lambda} C_n \times \Gamma_{\alpha_\lambda}$;
  3: $(a_3, \gamma_3) \xleftarrow{R} \mathbb{F}_q^{\alpha_\lambda} C_n \times \Gamma_{\alpha_\lambda}$;
  4: $pk_1 \leftarrow a_1 h \gamma_1$; $pk_2 \leftarrow a_2 h \gamma_2$;
  5: $k_0 \leftarrow a_2 pk_1 \widehat{\gamma}_2$; $k_1 \leftarrow a_3 h \gamma_3$;
  *and gives the triple $(pk_1, pk_2, k_b)$ to the adversary.*
- *The adversary outputs a bit $\widetilde{b} \in \{0, 1\}$*

*If $W_b$ is the event that $\mathcal{A}$ outputs 1 in experiment $b$, we define $\mathcal{A}$'s advantage in solving the Decisional Dihedral Product Problem for $\mathbb{F}_q^{\alpha_\lambda} D_{2n}$ as*

$$DDPadv[\mathcal{A}, \mathbb{F}_q^{\alpha_\lambda} D_{2n}] = |Pr[W_0] - Pr[W_1]|.$$

**Definition 4.7** (Decisional Dihedral Product Assumption). We say that the Decisional Dihedral Product (DDP) assumption holds for $\mathbb{F}_q^{\alpha_\lambda} D_{2n}$ if for all efficient adversaries $\mathcal{A}$ the quantity $\text{DDPadv}[\mathcal{A}, \mathbb{F}_q^{\alpha_\lambda} D_{2n}]$ is negligible.

Note that if an adversary wins the CDP Game with non-negligible probability for $\mathbb{F}_q^{\alpha_\lambda} D_{2n}$, then such adversary also wins the DDP Game with non-negligible probability for $\mathbb{F}_q^{\alpha_\lambda} D_{2n}$. Note that $\mathtt{h}$ is chosen as $\mathtt{h} = \mathtt{h}_1 + \mathtt{h}_2$, with $\mathtt{h}_1$ being a random non-zero element from $\mathbb{F}_q^{\alpha_\lambda} C_n$ and $\mathtt{h}_2$ being a random non-zero element from $\mathbb{F}_q^{\alpha_\lambda} C_n y$, to not let the attacker win the DDP Game trivially. Indeed if $\mathtt{h}$ is chosen as $\mathtt{h} = \mathtt{h}_1 + 0$ with $\mathtt{h}_1 \in \mathbb{F}_q^{\alpha_\lambda} C_{2n}$, then $\mathtt{k}_0 \in \mathbb{F}_q^{\alpha_\lambda} C_{2n}$ and $\mathtt{k}_1 \in \mathbb{F}_q^{\alpha_\lambda} C_{2n} y$ by Lemma 3.1. Similarly if $\mathtt{h}$ is chosen as $\mathtt{h} = 0 + \mathtt{h}_2$ with $\mathtt{h}_2 \in \mathbb{F}_q^{\alpha_\lambda} C_{2n} y$, then $\mathtt{k}_0 \in \mathbb{F}_q^{\alpha_\lambda} C_{2n} y$ and $\mathtt{k}_1 \in \mathbb{F}_q^{\alpha_\lambda} C_{2n}$ by Lemma 3.1. Therefore the attacker can win the DDP Game for both cases with non-negligible probability.

4.4. **Hardness of the DPD problem and related problems.** This section presents several algorithmic strategies an attacker may try to break the DPD Problem and hence any other related problems introduced in the previous section.

4.4.1. *Randomly Choosing.* The most simple strategy an adversary can try to win the DPD game with probability $1/|\mathbb{F}_q^{\alpha_\lambda} C_n \times \Gamma_{\alpha_\lambda}|$ is by simply choosing $(\widetilde{\mathtt{a}}, \widetilde{\gamma})$ from $\mathbb{F}_q^{\alpha_\lambda} C_n \times \Gamma_{\alpha_\lambda}$ at random. Therefore if $|\mathbb{F}_q^{\alpha_\lambda} C_n \times \Gamma_{\alpha_\lambda}|$ is not large enough so that $1/|\mathbb{F}_q^{\alpha_\lambda} C_n \times \Gamma_{\alpha_\lambda}|$ is not negligible, then the adversary will win the DPD game with non-negligible probability.

4.4.2. *Exhaustive Search Attack.* The adversary can also win the DPD game via an exhaustive search attack, i.e. by trying all possible $(\widetilde{\mathtt{a}}, \widetilde{\gamma})$ from $\mathbb{F}_q^{\alpha_\lambda} C_n \times \Gamma_{\alpha_\lambda}$ to find one that satisfies $\widetilde{\mathtt{a}} \mathtt{h} \widetilde{\gamma} = \mathtt{pk} = \mathtt{ah}\gamma$. However, its running time is linear in $|\mathbb{F}_q^{\alpha_\lambda} C_n \times \Gamma_{\alpha_\lambda}| = p^{nm} p^{m\lceil \frac{n+1}{2} \rceil} = p^{m(n+\lceil \frac{n+1}{2} \rceil)}$. Or equivalently the attacker may try enumerating bit strings of length $l = \lceil \log_2(p) m (n + \lceil \frac{n+1}{2} \rceil) \rceil$, with each representing a tuple $(\widetilde{\mathtt{a}}, \widetilde{\gamma}) \in \mathbb{F}_q^{\alpha_\lambda} C_n \times \Gamma_{\alpha_\lambda}$, to find one that satisfies $\widetilde{\mathtt{a}} \mathtt{h} \widetilde{\gamma} = \mathtt{pk}$. Therefore its running time is linear in $2^l$.

The attacker may improve the exhaustive search by running multiple parallel tasks as follows. Let $\mathcal{H} : \mathbb{F}_q^{\alpha_\lambda} D_{2n} \longrightarrow \{0, \ldots, q^{2n} - 1\}$ be a map defined by $\mathcal{H}(\sum_{g_i \in D_{2n}} \widetilde{a}_i \overline{g_i}) = \sum_{i=0}^{2n-1} \mathtt{rep}(\widetilde{a}_i) q^i$, where $\mathtt{rep}(\widetilde{a}_i)$ returns the non-negative integer used to represent $\widetilde{a}_i$, i.e. $\mathtt{rep}(\cdot)$ is a bijective map between the elements of $\mathbb{F}_q$ and the set $\{0, \ldots, q-1\}$. Therefore $\mathcal{H}$ is a bijection. Since $\mathcal{H}$ is a one-to-one map, then its inverse map $\mathcal{H}^{-1}$ exists and indeed can be easily constructed by converting a given base-10 integer $0 \leq a \leq q^{2n} - 1$ to a base-$q$ integer, hence obtaining a vector $[a_0, \ldots, a_{2n-1}]$, where $a_i$ are the integer representations of the field elements of $\mathbb{F}_q$. Therefore $\mathcal{H}^{-1}(a) = \sum_{g_i \in D_{2n}} \mathtt{rep}^{-1}(a_i) \overline{g_i}$.

Let us define a partition $\mathcal{P} = \{\mathcal{P}_0, \ldots, \mathcal{P}_{u-1}\}$ for $\mathcal{H}(\mathbb{F}_q^{\alpha_\lambda} C_n) = \{0, \ldots, q^n - 1\}$, then the attacker can create $u$ independent tasks $T_i$, with $i \in \{0, \ldots, u-1\}$, where each $T_i$ runs $\texttt{Search\_Over\_P}()$ with $\mathcal{P}_i, \mathtt{h}, \mathtt{pk}$ as parameters. Regarding $T_i$'s running time complexity, it is $\mathcal{O}(|\mathcal{P}_i \times \Gamma_{\alpha_\lambda}|)$.

```
1: function SEARCH_OVER_P(𝒫_k, h, pk)
2:     for a ∈ 𝒫_k do
3:         for γ ∈ Γ_{α_λ} do
4:             ã ← ℋ⁻¹(a);
5:             pk_c ← âhγ;
6:             if pk_c = pk then
```

```
 7:            return (ã, γ);
 8:         end if
 9:      end for
10:   end for
11: end function
```

4.4.3. *A space-time trade-off attack.* The attacker can also try a space-time trade-off algorithm to solve the DPD problem. Let $t$ be an integer in $\{0, \ldots, n\}$. Let us define

$$\mathcal{P}^t := \left\{ \sum_{i=0}^{t-1} \widetilde{a}_i \overline{x^i} \in \mathbb{F}_q^{\alpha_\lambda} C_n \right\},$$

with $\mathcal{P}^0 := \{0 \in \mathbb{F}_q^{\alpha_\lambda} C_n\}$. Similarly, let us define

$$\mathcal{P}_t := \left\{ \sum_{i=t}^{n-1} \widetilde{a}_i \overline{x^i} \in \mathbb{F}_q^{\alpha_\lambda} C_n \right\},$$

with $\mathcal{P}_n := \{0 \in \mathbb{F}_q^{\alpha_\lambda} C_n\}$. Therefore $\mathbb{F}_q^{\alpha_\lambda} C_n = \mathcal{P}^t \oplus \mathcal{P}_t$ as direct sum of $\mathbb{F}_q$-vector spaces. Additionally, let $\mathcal{T}$ be a hash table to store tuples $(\widetilde{\mathsf{a}}, \gamma) \in \mathcal{P}^t \times \Gamma_{\alpha_\lambda}$. The attack consists of two phases and proceeds as follows:

- The offline phase: The adversary runs the function Offline_Phase to store all tuples from $\mathcal{P}^t \times \Gamma_{\alpha_\lambda}$ in the table $\mathcal{T}$. Note this pre-computation can be done offline and only once.

```
1: function OFFLINE_PHASE(P^t, h)
2:    for ã₁ ∈ P^t do
3:       for γ ∈ Γ_{α_λ} do
4:          ã ← ã₁hγ;
5:          i ← H(ã);
6:          T[i].append((ã₁, γ));
7:       end for
8:    end for
9: end function
```

- The online phase: The adversary now runs the function Online_Phase to find a tuple $(\widetilde{\mathsf{a}}, \gamma) \in \mathbb{F}_q^{\alpha_\lambda} C_n \times \Gamma_{\alpha_\lambda}$, which satisfies $\widetilde{\mathsf{a}} \mathsf{h} \gamma = \mathsf{pk}$ by construction. Note that if the function Online_Phase finds a collision, i.e. a tuple $(\widetilde{\mathsf{a}_2}, \gamma) \in \mathcal{P}_t \times \Gamma_{\alpha_\lambda}$ such that the conditions (6) and (8) of the function Online_Phase are fulfilled, then $\widetilde{\mathsf{a}_1} \mathsf{h} \gamma = \mathsf{pk} - \widetilde{\mathsf{a}_2} \mathsf{h} \gamma$, where $(\widetilde{\mathsf{a}_1}, \gamma)$ is stored in table $\mathcal{T}$ at index $i$, and therefore $(\widetilde{\mathsf{a}_1} + \widetilde{\mathsf{a}_2}) \mathsf{h} \gamma = \mathsf{pk}$.

```
 1: function ONLINE_PHASE(P_t, h, pk)
 2:    for ã₂ ∈ P_t do
 3:       for γ ∈ Γ_{α_λ} do
 4:          ã ← pk − ã₂hγ;
 5:          i ← H(ã);
 6:          if T[i] is not empty then
 7:             for (ã₁, γ₁) ∈ T[i] do
 8:                if γ₁ = γ then
 9:                   return (ã₁ + ã₂, γ);
10:                end if
11:             end for
```

```
12:            end if
13:         end for
14:     end for
15:     return ⊥
16: end function
```

The function Offline_Phase stores $|\mathcal{P}^t \times \Gamma_{\alpha_\lambda}|$ tuples, so its running time and spacial complexity is $\mathcal{O}(|\mathcal{P}^t \times \Gamma_{\alpha_\lambda}|)$. Concerning the function Online_Phase, its running time is proportional to $|\mathcal{P}_t \times \Gamma_{\alpha_\lambda}|$. Since the adversary pre-computes and stores the table $\mathcal{T}$ by running the function Offline_Phase only once, then the attack's running time complexity is $\mathcal{O}(|\mathcal{P}_t \times \Gamma_{\alpha_\lambda}|)$.

4.4.4. *Grover's algorithm.* An adversary with access to a sufficiently large quantum computer may improve the exhaustive search attack by using the Grover's algorithm [11]. Recall that we can represent tuples $(\widetilde{\mathtt{a}}, \widetilde{\gamma})$ from $\mathbb{F}_q^{\alpha_\lambda} C_n \times \Gamma_{\alpha_\lambda}$ with bit-strings of length $l = \lceil \log_2(p) m(n + \lceil \frac{n+1}{2} \rceil) \rceil$. Given a quantum circuit for the testing function (the function T), Grover's algorithm shows that a binary representation $\mathtt{r}$ that satisfies the condition 3 of the function T can be found on a quantum computer in $\mathcal{O}(2^{l/2} T_1)$ steps, where $T_1$ is the time to evaluate $\mathtt{T}(\cdot)$. Therefore such adversary may find $\mathtt{r}$ in time proportional to $2^{l/2}$ using Grover's algorithm. One way to avoid this is to simply choose our key exchange's parameters such that $l/2$ is large enough so that this search be infeasible.

```
1: function T(r ∈ {0, 1}^l, h, pk)
2:     (ã, γ̃) ← getPair(r);
3:     if ãhγ̃ = pk then
4:         return 1;
5:     end if
6:     return 0;
7: end function
```

4.4.5. *Shor's algorithm and variants.* Given an $\mathtt{h}$ in $F^{\alpha_\lambda} D_{2n}$, the set $\{\mathtt{a}\mathtt{h}\gamma | (\mathtt{a}, \gamma) \in \mathbb{F}_q^{\alpha_\lambda} C_n \times \Gamma_{\alpha_\lambda}\}$ is not even a semi-group under the twisted algebra multiplication. So it is not clear how a period-finding algorithm may be exploited in this setting. Hence Shor's algorithm variants may not apply to solving the DDP problem either [20, 21].

4.4.6. *Algebraic perspective.* From an algebraic viewpoint, the authors of [8, 19] introduces an attack to break some constructions using matrices over group rings when the characteristic of $\mathbb{F}_q$, $p$, does not divide the order of $G$. This attack is based on Maschke's Theorem that asserts that a group algebra $\mathbb{F}_q G$ is semi-simple if and only the characteristic of $\mathbb{F}_q$ does not divide the order of $G$ (Satz 7.17 [23]). Since Maschke's Theorem is also valid for twisted group algebras with a 2-cocycle defined over a finite field, a similar attack might break the DPD problem. However, that attack can no longer be applied to our construction because the characteristic of $\mathbb{F}_q$ is chosen to divide $|D_{2n}|$, and thus $\mathbb{F}_q^{\alpha_\lambda} D_{2n}$ is not semi-simple.

4.5. **Security analysis in the authenticated-links adversarial model.** This subsection is devoted to analysing further our key exchange protocol in a appropriate security model [1, 13, 3]. In particular, we aim at proving that our protocol is session-key secure in the authenticated-links adversarial model (AM) of Canetti and Krawczyk [3], assuming the DDP assumption holds for $\mathbb{F}_q^{\alpha_\lambda}$. We first recall the definition of session-key security in the authenticated-links adversarial model of Canetti and Krawczyk [3].

Let $P = \{P_1, P_2, \ldots, P_n\}$ be a finite set of parties and let $\mathcal{A}$ be an adversary that controls all communication, but $\mathcal{A}$ cannot inject or modify messages, except for messages sent by corrupted parties or sessions. Additionally, $\mathcal{A}$ may decide not to deliver a message at all, but if $\mathcal{A}$ decides to deliver a message $m$, $\mathcal{A}$ can do so to the proper destination for $m$, only once and without modifying $m$.

Parties give outgoing messages to $\mathcal{A}$, who has control over their delivery via the `Send` query. $\mathcal{A}$ can activate a party $P_i$ by `Send` queries, i.e. the adversary has control over the creation of protocol sessions, which take place within each party. Two sessions $s_1$ and $s_0$ are matching if the outgoing messages of one are the incoming messages of the other, and vice versa. Additionaly, $\mathcal{A}$ is allowed to query the oracles `SessionStateReveal`, `SessionKeyReveal`, and `Corrupt`.

- If $\mathcal{A}$ query the `SessionStateReveal` oracle for a specified session-id $s$ within some party $P_i$, then $\mathcal{A}$ obtains the contents of the specified session-id $s$ within $P_i$, including any secret information. This event is noted and hence produces no further output.
- If $\mathcal{A}$ query the `SessionKeyReveal` for a specified session-id $s$, then $\mathcal{A}$ obtains the session key for the specified session $s$, assuming that $s$ has an associated session.
- If $\mathcal{A}$ query the `Corrupt` oracle for a specified party $P_i$, then $\mathcal{A}$ takes over the party $P_i$, i.e. $\mathcal{A}$ has access to all information in $P_i$'s memory, including long-lived keys and any session-specific information still stored. A corrupted party produces no further output.

Additionally, $\mathcal{A}$ is given access to the `test` oracle, which can be queried once and at any stage to a completed, fresh, unexpired session-id $s$. On input $s$, the `test` oracle chooses $b \xleftarrow{R} \{0, 1\}$, then it outputs the session key for the specified session-id $s$ if $b = 0$. Otherwise, it returns a random value in the key space. Also, $\mathcal{A}$ can issue subsequent queries as desired, with the exception that it cannot expose the test session. At any point, the adversary can try to guess $b$. Let $\text{Guess}[\mathcal{A}, \mathbb{F}_q^{\alpha_\lambda} D_{2n}]$ be the event that $\mathcal{A}$ correctly guesses $b$, and define the advantage $\text{SKAdv}[\mathcal{A}, \mathbb{F}_q^{\alpha_\lambda} D_{2n}] = |\text{Guess}[\mathcal{A}, \mathbb{F}_q^{\alpha_\lambda} D_{2n}] - 1/2|$.

**Theorem 4.8.** *If the DDP assumption holds for $\mathbb{F}_q^{\alpha_\lambda} D_{2n}$, then our key exchange protocol is session-key secure in the the authenticated-links adversarial model, i.e. for any $\mathcal{A}$ in the authenticated-links adversarial model (AM), then the following holds*

1. *The key-exchange protocol satisfies the property that if two uncorrupted parties complete matching sessions, then they both output the same key.*
2. *$SKAdv[\mathcal{A}, \mathbb{F}_q^{\alpha_\lambda} D_{2n}]$ is negligible.*

*Proof.* The proof is an adaptation to that one given by Canetti and Krawczyk [3] for the Diffie-Hellman Protocol over $\mathbb{Z}_q^*$. To prove the first item, let us suppose that if $P_i$ and $P_j$ are uncorrupted during the exchange of the key and both complete the protocol for session-id $s$, then they both establish the same key. Indeed, since the choice of $\alpha_\lambda$, we have $\mathtt{a}_i \mathtt{a}_j = \mathtt{a}_j \mathtt{a}_i$ and $\gamma_i \widehat{\gamma_j} = \gamma_j \widehat{\gamma_i}$, then

$$\mathtt{k}_i = \mathtt{a}_i \mathtt{pk}_j \widehat{\gamma_i} = \mathtt{a}_i \mathtt{a}_j \mathtt{h} \gamma_j \widehat{\gamma_i} = \mathtt{a}_j \mathtt{a}_i \mathtt{h} \gamma_i \widehat{\gamma_j} = \mathtt{a}_j \mathtt{pk}_i \widehat{\gamma_j} = \mathtt{k}_j$$

To prove the second item. We proceed by assuming to the contrary, i.e. there is an adversary $\mathcal{A}$ in the authentication-links model against our protocol that has a non-negligible advantage $\epsilon$ in guessing the bit $b$ chosen by the test oracle (when queried). Let $l$ be an upper bound on the number of sessions invoked by $\mathcal{A}$ in any interaction. We now construct a distinguisher $\mathcal{D}$ for the DDP problem as shown next.

1: **function** $\mathcal{D}(h, \mathbb{F}_q^{\alpha_\lambda} D_{2n}, \text{pk}_1, \text{pk}_2, \text{k})$
2:     $r \xleftarrow{R} \{1, \ldots, l\}$.
3:     Invoke $\mathcal{A}$ on a simulated interaction in the AM with parties $P_1, \ldots, P_n$, except for the $r_{th}$ session.
4:     For the $r$-th session, let $P_i$ send $(P_i, s, \text{pk}_i = \mathtt{a}_i \mathtt{h} \gamma_i)$ to $P_j$ and let $P_j$ send $(P_j, s, \text{pk}_j = \mathtt{a}_j \mathtt{h} \gamma_j)$ to $P_i$.
5:     **if** $r$-th session is chosen by $\mathcal{A}$ as the test session **then**
6:         give $\mathtt{k}$ to $\mathcal{A}$ as the answer to his query.
7:         $d \leftarrow \mathcal{A}(\mathtt{k})$
8:     **else**
9:         $d \xleftarrow{R} \{0, 1\}$.
10:     **end if**
11:     **return** $d$
12: **end function**

Suppose that $\mathcal{A}$ chooses $r$-th as the test session. Since $\mathtt{k}_0$ or $\mathtt{k}_1$ is given to $\mathcal{D}$ by its DDP challenger, thus $\mathcal{A}$ receives either of the two keys. Therefore, the probability that $\mathcal{A}$ distinguish correctly is $1/2 + \epsilon$ with non-negligible $\epsilon$ (by assumption).

Suppose that $\mathcal{A}$ does not choose $r$-th as the test session, then $\mathcal{D}$ always output a random bit, and then the probability to distinguish the input correctly is $1/2$.

Note that the probability that the test-session and the $r$-th session coincide is $1/l$, and that they do not coincide is $1 - 1/l$, therefore the overall probability for $\mathcal{D}$ to win the DDP Game is $1/(2l) + \epsilon/l + 1/2 - 1/(2l) = 1/2 + \epsilon/l$, which is non-negligible. □

## 5. A Probabilistic Encryption Scheme.

We now present a probabilistic public key encryption scheme following a similar paradigm as the one followed by [9] and other recent post-quantum constructions [4, 18]. Let $\mathbb{F}_q$ be a finite field with $q = p^m$ and $p > 2$, $D_{2n}$ be the dihedral group of order $2n$, $\alpha_\lambda$ be the 2-cocycle from Lemma 3.4 for a non-square $\lambda \in \mathbb{F}_q$, and $\mathtt{h} = \mathtt{h}_1 + \mathtt{h}_2$ be a public element in $\mathbb{F}_q^{\alpha_\lambda} D_{2n}$, where $\mathtt{h}_1$ is a random non-zero element from $\mathbb{F}_q^{\alpha} C_n$ and $\mathtt{h}_2$ is a random non-zero element from $\mathbb{F}_q^{\alpha_\lambda} C_n y$.

Let $\mathcal{SK} = \mathbb{F}_q^{\alpha_\lambda} C_{2n} \times \Gamma_{\alpha_\lambda}$ be the secret key space, $\mathcal{M} = \mathbb{F}_q^{\alpha_\lambda} D_{2n}$ be the message space, $\mathcal{C} = \mathbb{F}_q^{\alpha_\lambda} D_{2n} \times \mathbb{F}_q^{\alpha_\lambda} D_{2n}$ be the ciphertext space and $\mathcal{PK} = \mathbb{F}_q^{\alpha_\lambda} D_{2n}$ be the public key space. We next show the public-key encryption scheme $\mathcal{E} = (\text{Gen}, \text{Enc}, \text{Dec})$ derived from the previous key exchange protocol.

1: **function** GEN($\mathtt{h} \in \mathbb{F}_q^{\alpha_\lambda} D_{2n}$)
2:     $(\mathtt{a}_1, \gamma_1) \xleftarrow{R} \mathcal{SK}$;
3:     $\text{pk} \leftarrow \mathtt{a}_1 \mathtt{h} \gamma_1$;
4:     $\text{sk} \leftarrow (\mathtt{a}_1, \gamma_1)$;
5:     **return** $\text{pk}, \text{sk}$;
6: **end function**

1: **function** DEC($\mathtt{c} \in \mathcal{C}, \text{sk} \in \mathcal{SK}$)
2:     $(\mathtt{a}_1, \gamma_1) \leftarrow \text{sk}$;
3:     $(\mathtt{c}_1, \mathtt{c}_2) \leftarrow \mathtt{c}$;
4:     $\mathtt{k} \leftarrow \mathtt{a}_1 \mathtt{c}_1 \widehat{\gamma}_1$;
5:     $\mathtt{m} \leftarrow \mathtt{c}_2 - \mathtt{k}$;
6:     **return** $\mathtt{m}$;
7: **end function**

```
1: function Enc(m ∈ M, pk ∈ PK, r₂ ∈     4:     c₂ ← m + a₂pk γ̂₂;
   SK, h ∈ F_q^{αλ} D_{2n})                5:     c ← (c₁, c₂);
2:     (a₂, γ₂) ← r₂;                      6:     return c;
3:     c₁ ← a₂hγ₂;                         7: end function
```

**Lemma 5.1** (Correctness). *Let $h$ be a public element in $\mathbb{F}_q^{\alpha\lambda} D_{2n}$, $m \in \mathcal{M}$ a message, $r_2 \xleftarrow{R} \mathcal{SK}$ a random secret key and*

$$(pk, sk) \leftarrow \text{Gen}(h),$$

*then*

$$m \leftarrow \text{Dec}(\text{Enc}(m, pk, r_2, h), sk)$$

*Proof.* Since

$$(c_1 = a_2 h \gamma_2, c_2 = m + a_2 pk \widehat{\gamma}_2) \leftarrow \text{Enc}(m, pk, r_2, h)$$

and $sk = (a_1, \gamma_1)$, then

$$k = a_1 c_1 \widehat{\gamma}_1 = a_1 a_2 h \gamma_2 \widehat{\gamma}_1 = a_2 a_1 h \gamma_1 \widehat{\gamma}_2 = a_2 pk \widehat{\gamma}_2,$$

and therefore

$$c_2 - k = m + a_2 pk \widehat{\gamma}_2 - a_2 pk \widehat{\gamma}_2 = m$$

□

Since our public-key encryption scheme $\mathcal{E}$ is based on the previous key exchange protocol, then $\mathcal{E}$ does not provide message confidentiality if the CDP assumption does not hold for $\mathbb{F}^{\alpha\lambda} D_{2n}$. We now prove that $\mathcal{E}$ is semantically secure under the stronger DDP assumption.

**Theorem 5.2.** *If the DDP assumption holds for $\mathbb{F}_q^{\alpha\lambda} D_{2n}$, then $\mathcal{E}$ is semantically secure.*

*Proof.* Let $\mathcal{A}$ be an efficient adversary. Let us define Game 0 to be the semantic security (SS) attack game against $\mathcal{A}$.

```
1: function GAME0()
2:     (pk₁, (a₁, γ₁)) ← Gen(h);
3:     (m₀, m₁) ← A(pk₁);
4:     b ←R {0, 1}; (a₂, γ₂) ←R SK;
5:     pk₂ ← a₂hγ₂; k ← a₂pk₁γ̂₂;
6:     c ← m_b + k;
7:     b̃ ← A(pk₁, pk₂, c);
8:     return [[b = b̃]];
9: end function
```

In the above game and subsequent games $[[b = \widetilde{b}]]$ returns 1 if both $b$ and $\widetilde{b}$ are equals, or 0 otherwise. Additionally, note that $\mathcal{A}$ presents the challenger with two messages $m_0, m_1 \in \mathcal{M}$ with both having the same length in bits. We now define $S_0$ to be the event that Game 0 returns 1, then $\mathcal{A}$'s SS-advantage is $|\Pr[S_0] - 1/2|$. We will prove $|\Pr[S_0] - 1/2|$ is negligible under the DDP assumption. We now define Game 1 by making some changes to Game 0 as follows.

```
1: function GAME1()
2:     (pk₁, (a₁, γ₁)) ← Gen(h);
3:     (m₀, m₁) ← A(pk₁);
```

```
 4:      b ←R {0,1}; (a₂, γ₂) ←R SK;
 5:      pk₂ ← a₂hγ₂;
 6:      (a₃, γ₃) ←R SK; k ← a₃hγ₃;
 7:      c ← m_b + k;
 8:      b̃ ← A(pk₁, pk₂, c);
 9:      return [[b = b̃]]
10: end function
```

Let $S_1$ be the event that Game 1 returns 1. Since $\mathtt{b}, \mathtt{pk}_1, \mathtt{pk}_2, \mathtt{c}$ are mutually independent, then it follows that $\mathtt{b}$ and $\widetilde{\mathtt{b}} \leftarrow \mathcal{A}(\mathtt{pk}_1, \mathtt{pk}_2, \mathtt{c})$ are independent. Therefore $\Pr[S_1] = 1/2$.

We now construct an adversary $\mathcal{B}$ that plays the DDP Attack Game 3. In particular this adversary $\mathcal{B}$ plays the role of challenger to $\mathcal{A}$. It obtains the triple $(\mathtt{pk}_1, \mathtt{pk}_2, \mathtt{k})$ from its own challenger, and then sends $\mathtt{pk}_1$ to $\mathcal{A}$; when receives $(\mathtt{m}_0, \mathtt{m}_1)$ from $\mathcal{A}$, it selects a random bit $\mathtt{b}$, encrypts $\mathtt{m}_\mathtt{b}$ with $\mathtt{k}$ and sends $\mathtt{pk}_1, \mathtt{pk}_2, \mathtt{c}$ to $\mathcal{A}$. It finally obtains a response bit $\widetilde{\mathtt{b}}$ from $\mathcal{A}$ and outputs 1 if $\widetilde{\mathtt{b}} = \mathtt{b}$, or 0 otherwise. Note that since $\mathcal{A}$ is an efficient adversary, so is $\mathcal{B}$. The logic of $\mathcal{B}$'s challenger and $\mathcal{B}$ are shown next.

```
1: function GAMEDDP(b)                    1: function B(pk₁, pk₂, k)
2:     (a₁, γ₁) ←R SK;                   2:     (m₀, m₁) ← A(pk₁);
3:     (a₂, γ₂) ←R SK;                   3:     b ←R {0, 1};
4:     (a₃, γ₃) ←R SK;                   4:     c ← m_b + k;
5:     pk₁ ← a₁hγ₁; pk₂ ← a₂hγ₂;          5:     b̃ ← A(pk₁, pk₂, c);
6:     k₀ ← a₂pk₁γ̂₂; k₁ ← a₃hγ₃;          6:     return [[b = b̃]]
7:     b̃ ← B(pk₁, pk₂, k_b);              7: end function
8:     return [[b̃ = 1]];
9: end function
```

Recall $W_\mathtt{b}$ is the event that $\mathcal{B}$ outputs 1 in Game DDP(b) (i.e. it returns 1), so the $\mathcal{B}$'s advantage in solving the Decisional Dihedral Product Problem for $\mathbb{F}_q^{\alpha_\lambda} D_{2n}$ is given by

$$\mathrm{DDPadv}[\mathcal{B}, \mathbb{F}_q^{\alpha_\lambda} D_{2n}] := |\Pr[W_0] - \Pr[W_1]|.$$

Note that when $\mathcal{B}$'s challenger is playing experiment $\mathtt{b} = 0$ of the Game DDP, $\mathcal{A}$ is essentially playing Game 0, since $\mathcal{B}$ receives from its challenger $\mathtt{pk}_1 = \mathtt{a}_1\mathtt{h}\gamma_1$, $\mathtt{pk}_2 = \mathtt{a}_2\mathtt{h}\gamma_2$, $\mathtt{k} = \mathtt{a}_2\mathtt{pk}_1\widehat{\gamma}_2$. Therefore $\Pr[W_0] = \Pr[S_0]$. Analogously, when $\mathcal{B}$'s challenger is playing experiment $\mathtt{b} = 1$ of the Game DDP, $\mathcal{A}$ is essentially playing Game 1, since $\mathcal{B}$ receives from its challenger $\mathtt{pk}_1 = \mathtt{a}_1\mathtt{h}\gamma_1$, $\mathtt{pk}_2 = \mathtt{a}_2\mathtt{h}\gamma_2$, $k = \mathtt{a}_3\mathtt{h}\gamma_3$. Therefore $\Pr[W_1] = \Pr[S_1]$. Since $|\Pr[W_0] - \Pr[W_1]|$ is negligible by hypothesis, then we have that $|\Pr[S_0] - 1/2|$ is negligible and the assertion follows. □

It is well known that if a public-key encryption scheme is semantically secure, then it is also semantically secure against chosen-plaintext attack (IND-CPA) [2]. Therefore $\mathcal{E}$ is CPA-secure under the DDP assumption.

6. **A Key Encapsulation Mechanism.** In this section, we present a CCA-secure key encapsulation mechanism by applying a generic transformation of Hofheinz, Hövelmanns, and Kiltz [12] to $\mathcal{E}$. Let $\mathcal{K} = \{0,1\}^{l_1}$ be the keyspace and $\mathtt{rep}(x)$ be a function that simply returns the binary representation of $x$. Additionally, we construct two hash functions:

- The function $\mathcal{G}_1 : \{0,1\}^* \longrightarrow \mathcal{SK}$ takes a bit-string, say x, as input and then apply a suitable cryptographic hash function like $\text{SHAKE}_{256}$ to it. In the notation of [7], $\mathcal{G}_1(\text{x}) = \text{SHAKE}_{256}(\text{x}, \text{o})$, where $\text{o} = \lceil \log_2(p) \rceil m(n + \lceil \frac{n+1}{2} \rceil)$ is the bit length of the output. From this bit-string, the corresponding pair $(\text{a}, \gamma) \in \mathcal{SK}$ can be obtained easily.
- Similarly, the function $\mathcal{G}_2 : \{0,1\}^* \longrightarrow \mathcal{K}$ is constructed by applying $\text{SHAKE}_{256}$ to the input. That is, $\mathcal{G}_2(\text{x}) = \text{SHAKE}_{256}(\text{p}_1 || \text{x}, l_1)$, where $\text{p}_1$ is a prepended fixed bit-string to make it different from $\mathcal{G}_1$.

Applying the generic transformation $\text{U}^{\not\perp}[\text{T}[\mathcal{E}, \mathcal{G}_2], \mathcal{G}_1]$ from [12], we get

$$\text{KEM} = (\text{KeyGen}, \text{Encaps}, \text{Decaps}),$$

where

```
1: function KeyGen(h)
2:     (pk, sk) ← Gen(h);
3:     s ←R M;
4:     return (s, sk, pk);
5: end function

1: function Encaps(pk, h)
2:     m ←R M;
3:     r ← G1(rep(m)||rep(pk));
4:     c ← Enc(m, pk, r, h);
5:     K ← G2(rep(m)||rep(c));
6:     return (c, K);
7: end function
```

```
1: function Decaps((s, pk, sk), c, h)
2:     m ← Dec(c, sk);
3:     r ← G1(rep(m)||rep(pk));
4:     if c = Enc(m, pk, r, h) then
5:         K ← G2(rep(m)||rep(c));
6:         return K;
7:     else
8:         return G2(rep(s)||rep(c));
9:     end if
10: end function
```

6.1. **Implementation.** We implemented our proposed public-key encryption scheme and key encapsulation mechanism as a proof-of-concept in Python. The interested reader can see it on Google Colaboratory [6].

6.1.1. *Dihedral Group.* To implement a dihedral group of order $2n$, we simply represent a diehdral group element $g = x^{i_1} y^{j_1}$ as the integer $k_1 = j_1 \cdot n + i_1$. Also, we compute a $2n \times 2n$ integer array table such that the row $\text{table}[k_1]$, $0 \leq k_1 < 2n$, stores a $2n$ array with the integer representations of $g, gx, gx^2, \ldots gx^{n-1}, gy, \ldots, gx^{n-1}y$. To compute the operation of two given group elements $g = x^{i_1} y^{j_1}$ and $h = x^{i_2} y^{j_2}$, we simply return $\text{table}[k_1][k_2]$, where $k_1 = j_1 \cdot n + i_1$ and $k_2 = j_2 \cdot n + i_2$. To compute the multiplicative inverse of a given group element $g = x^{i_1} y^{j_1}$, the function $\text{inverse}(k_1)$ returns 0 if $k_1 = 0$, or $n - k_1$ if $1 \leq k_1 < n$, or $k_1$ if $n \leq k_1 < 2n$.

6.1.2. *2-cocycle $\alpha_\lambda$.* The 2-cocycle $\alpha_\lambda$ is implemented efficiently as described next. Given $k_1$ and $k_2$, two representations of two group elements, then the function $\text{cocycle}(k_1, k_2)$ returns $\lambda \in \mathbb{F}_q$ if $n \leq k_1 < 2n$ and $n \leq k_2 < 2n$. Otherwise, it returns $1 \in \mathbb{F}_q$. As a consequence of Lemma 3.6, $\lambda$ may be chosen as a non-square in $\mathbb{F}_q$. The following function describes exactly how $\lambda$ is chosen:

```
1: function getLambda()
2:     λ ← getRandomFieldElement()
3:     r ← λ^((q-1)/2)
4:     while r = 1 do
5:         λ ← getRandomFieldElement()
```

6:         $\mathtt{r} \leftarrow \lambda^{(q-1)/2}$
7:     **end while**
8:     **return** $\lambda$
9: **end function**

The function `getRandomFieldElement()` returns a field element chosen uniformly randomly in $\mathbb{F}_q$.

6.1.3. *Twisted algebra* $\mathbb{F}_q^{\alpha_\lambda} D_{2n}$. A element $a = \sum_{i=0}^{n-1} a_i \overline{x^i} + \sum_{i=0}^{n-1} a_{n+i} \overline{x^i y}$ in the algebra $\mathbb{F}_q^{\alpha_\lambda} D_{2n}$ is represented as a array of $2n$ field elements $\mathtt{a} = [\mathtt{a}_0, \mathtt{a}_1, \mathtt{a}_2, \ldots, \mathtt{a}_{2n-1}]$, where $\mathtt{a}_i$ is the representation of the field element $a_i \in \mathbb{F}_q$. Therefore the addition and product of two elements of this algebra is easily implemented as shown next.

1: **function** ADDITION($\mathtt{a}, \mathtt{b}$)
2:     $\mathtt{c} \leftarrow [0, \cdots, 0]$
3:     **for** $(i \leftarrow 0; i < 2n; i \leftarrow i+1)$ **do**
4:         $\mathtt{c}[i] \leftarrow \mathtt{a}[i] + \mathtt{b}[i];$
5:     **end for**
6:     **return** $\mathtt{c}$
7: **end function**

1: **function** PRODUCT($\mathtt{a}, \mathtt{b}$)
2:     $\mathtt{c} \leftarrow [0, \cdots, 0]$
3:     **for** $(i \leftarrow 0; i < 2n; i \leftarrow i+1)$ **do**
4:         **for** $(j \leftarrow 0; j < 2n; j \leftarrow j+1)$ **do**
5:             $k \leftarrow \mathtt{table}[i, j];$
6:             $\mathtt{fe} \leftarrow \mathtt{a}[i] \cdot \mathtt{b}[j] \cdot \mathtt{cocycle}(i, j);$
7:             $\mathtt{c}[k] \leftarrow \mathtt{c}[k] + \mathtt{fe};$
8:         **end for**
9:     **end for**
10:     **return** $\mathtt{c}$
11: **end function**

Note that the addition function has a cost of $2n$ field additions to compute an algebra element $\mathtt{c}$. Regarding the product of two algebra elements, it has a cost of $4n^2$ field additions and $8n^2$ field multiplications. Also, we implement the function `adjunct`, which computes the adjunct of a algebra element and its cost is $2n$ multiplications, and also implement functions for computing a random element in $\Gamma_{\alpha_\lambda}$ ( $\mathbb{F}_q^{\alpha_\lambda} D_{2n}$, $\mathbb{F}_q^{\alpha_\lambda} C_n$ and $\mathbb{F}_q^{\alpha_\lambda} C_n y$).

1: **function** ADJUNCT($\mathtt{a}$)
2:     $\mathtt{c} \leftarrow [0, \cdots, 0]$
3:     **for** $(i \leftarrow 0; i < 2n; i \leftarrow i+1)$ **do**
4:         $j \leftarrow \mathtt{inverse}(i)$
5:         $\mathtt{c}[j] = \mathtt{a}[i] \cdot \mathtt{cocycle}(i, j)$
6:     **end for**
7:     **return** $\mathtt{c}$
8: **end function**

1: **function** GETRANDOMFROMT()
2:     $\mathtt{c} \leftarrow [0, \cdots, 0]$
3:     $\mathtt{c}[n] \leftarrow \mathtt{getRandomFieldElement}()$
4:     $n_1 \leftarrow n/2$
5:     **for** $(i \leftarrow 1; i \leq n_1; i \leftarrow i+1)$ **do**
6:         $\mathtt{c}[i+n] \leftarrow \mathtt{getRandomFieldElement}()$
7:         $\mathtt{c}[n + (n-i) \bmod n] \leftarrow \mathtt{c}[i+n]$
8:     **end for**
9:     **return** $\mathtt{c}$
10: **end function**

1: **function** GETRANDOMFD2N()
2:     $\mathtt{c} \leftarrow [0, \cdots, 0]$
3:     **for** $(i \leftarrow 0; i < 2n; i \leftarrow i+1)$ **do**
4:         $\mathtt{c}[i] \leftarrow \mathtt{getRandomFieldElement}()$
5:     **end for**
6:     **return** $\mathtt{c}$
7: **end function**

1: **function** GETRANDOMFCN()
2:     $\mathtt{c} \leftarrow [0, \cdots, 0]$
3:     **for** $(i \leftarrow 0; i < n; i \leftarrow i+1)$ **do**
4:         $\mathtt{c}[i] \leftarrow \mathtt{getRandomFieldElement}()$
5:     **end for**
6:     **return** $\mathtt{c}$
7: **end function**

1: **function** GETRANDOMFCNY()
2:     $\mathtt{c} \leftarrow [0, \cdots, 0]$
3:     **for** $(i \leftarrow n; i < 2n; i \leftarrow i+1)$ **do**
4:         $\mathtt{c}[i] \leftarrow \mathtt{getRandomFieldElement}()$
5:     **end for**

```
 6:        return c                              7: end function
```

Finally, since $h$ should be chosen as stated in Section 4.3, we also implement the following function to compute a random public element $h$.

```
 1: function GETPUBLICELEMENT()
 2:     sw₁ ← False
 3:     while not sw₁ do
 4:         a ← getRandomFD2n()
 5:         i ← 0
 6:         sw₂ ← False
 7:         while i < n and not sw₂ do
 8:             if a[i] ≠ 0 then
 9:                 sw₂ ← True
10:             end if
11:             i ← i + 1
12:         end while
13:         i ← n
14:         sw₃ ← False
15:         while i < 2n and not sw₃ do
16:             if a[i] ≠ 0 then
17:                 sw₃ ← True
18:             end if
19:             i ← i + 1
20:         end while
21:         sw₁ ← sw₂ and sw₃
22:     end while
23:     return a
24: end function
```

6.1.4. *Parameters choice.* For our KEM, we propose to use the parameters shown by Table 1, which provide varying degrees of security.

| $p$ | $m$ | $n$ | $l_1$ (bits) | $l$ (bits) |
|---|---|---|---|---|
| 19 | 1 | 19 | $\{128, 192, 256\}$ | 124 |
| 23 | 1 | 23 | $\{128, 192, 256\}$ | 149 |
| 31 | 1 | 31 | $\{128, 192, 256\}$ | 200 |
| 41 | 1 | 41 | $\{128, 192, 256\}$ | 264 |

TABLE 1. Proposed parameters

Table 1 shows four sets of parameters providing various degrees of security. The value of $l_1 \in \{128, 192, 256\}$ refers to the length of the output key and $l$ refers to the value we introduced in Section 4.4.4. To see the code of our implementation, please see [6].

7. **Conclusions.** In this paper, we presented a key encapsulation mechanism derived from a probabilistic public-key encryption scheme based on a twisted dihedral algebra. We think this work can be extended to other twisted group algebras for other families of groups, such as metacyclic groups. On the other hand, we think this work may be extended to twisted group rings, i.e. on other finite rings that are not necessarily fields.